\documentstyle[12pt,amsfonts]{article}
\parskip=\medskipamount

\newcommand {\cD}{{\cal D}}

\newcommand {\cF}{{\cal F}}
\newcommand {\cG}{{\cal G}}

\newcommand {\cL}{{\cal L}}

\newcommand {\cN}{{\cal N}}

\newcommand {\cV}{{\cal V}}
\newcommand {\cW}{{\cal W}}

\def\a{\alpha}
\def \bi{\bibitem}

\def\b{\beta}

\def\d{\delta}

\def\f{\phi}
\def\g{\gamma}

\def\j{\psi}
\def\k{\kappa}
\def\l{\lambda}
\def\m{\mu}

\def\o{\omega}
\def\p{\pi}
\def\q{\theta}
\def\r{\rho}
\def\s{\sigma}

\def\z{\zeta}

\def\F{\Phi}
\def\J{\Psi}
\def\L{\Lambda}
\def\O{\Omega}

\def\S{\Sigma}
\def\U{\Upsilon}

\newcommand{\ad}{{\dot{\alpha}}}                           
\newcommand{\bd}{{\dot{\beta}}}                            
\newcommand{\ve}{\varepsilon}                            
\newcommand{\cDB}{{\bar \cD}}                            
\newcommand{\gd}{{\dot{\gamma}}}

\newcommand{\pa}{\partial}                           
\newcommand{\hf}{\frac12}

\newcommand{\sect}[1]{\setcounter{equation}{0}\section{#1}}

\newcommand{\be}{\begin{equation}}
\newcommand{\ee}{\end{equation}}
\newcommand{\bea}{\begin{eqnarray}}
\newcommand{\eea}{\end{eqnarray}}
\newcommand{\non}{\nonumber}

\def \intss{\int\!\!{\rm d}^8z}

\def \Dsqac{(\cDB^2 - 4 R)}
\begin{document}

\begin{titlepage}

\begin{flushright}
hep-th/0412190\\
December, 2004
\end{flushright}
\vspace{5mm}

\begin{center}
{\Large\bf  
On massive tensor multiplets
}
\end{center}

\begin{center}

{\large Sergei M. Kuzenko}
\vspace{2mm}

\footnotesize{
{\it School of Physics, The University of Western Australia,\\
35 Stirling Highway, Crawley W.A. 6009, Australia}} \\
{\tt kuzenko@cyllene.uwa.edu.au}
\vspace{2mm}

\end{center}
\vspace{5mm}

\begin{abstract}
\baselineskip=14pt
\noindent
Massive tensor multiplets have
recently been scrutinized in  hep-th/0410051 and
hep-th/0410149, as they appear in orientifold
compactifications of type IIB string theory.
Here we formulate several dually equivalent 
models for massive $\cN=1, \, 2$ tensor multiplets 
in four space-time dimensions.  
In the $\cN=2$ case, we employ  harmonic
and projective superspace techniques.
\end{abstract}

\vfill
\end{titlepage}

\newpage
\setcounter{page}{1}
\renewcommand{\thefootnote}{\arabic{footnote}}
\setcounter{footnote}{0}

\sect{Introduction}
Recently, there has been renewed interest 
in 4D $\cN=1$ massive tensor multiplets 
and their couplings to scalar and vector 
multiplets \cite{DF,LS}.
Such  interest is primarily motivated by  
the fact that massive two-forms 
naturally appear in four-dimensional  
$\cN=2$ supergravity  theories 
obtained from (or related to) compactifications 
of type II string theory on Calabi-Yau threefolds
in the presence of  both electric and magnetic 
fluxes \cite{LM,DSV}. This clearly provides enough 
ground for undertaking a more detailed study of 
massive $\cN=1$ and $\cN=2$ tensor 
multiplets. 

In $\cN=1$ supersymmetry, 
the massive tensor multiplet 
(as a dual version of the massive vector multiplet)
was introduced twenty five years ago \cite{Siegel}, 
and since then 
this construction\footnote{The work of \cite{Siegel}
is actually more famous for the massless tensor 
multiplet (as a dual version of the chiral scalar multiplet)  
introduced in it, see also \cite{Gates,LR}.}
has been reviewed in 
two textbooks  \cite{GGRS,BK}.
In the original formulation \cite{Siegel}, 
the mass parameter, $m^2$, 
in the action 
\be
\label{or-massive-tensor}
S_{\rm tensor} = - \hf \intss\, G^2
- \hf \left\{  m^2 \int {\rm d}^6z \, \j^\a \j_\a
+{\rm c.c.} \right\}
\ee
was chosen to be real. Here 
\be
G = \hf( D^\a \j_\a + {\bar D}_\ad {\bar \j}^\ad)~, 
\qquad  \quad {\bar D}_\ad \j_\a =0~, 
\label{ten-field-strength}
\ee
where the dynamical variable $\j_\a$
is an arbitrary chiral spinor superfield,
and the (massless gauge) 
field strength $G$ is a real linear superfield, 
$D^2 G = {\bar D}^2 G=0$.
The choice of a real  mass parameter seemed 
to be natural  in the sense that, for $M =m$, 
the above system is dual to 
the massive vector multiplet model\footnote{This
duality is a supersymmetric version of the duality 
between massive one- and two-forms 
\cite{massive}.}  
\be
\label{massive-vector}
S_{\rm vector} = {1\over 4}  \int {\rm d}^6z \,
W^\a W_\a 
+ \hf M^2 \int {\rm d}^8z \, V^2~,
\qquad W_\a = -{ 1 \over 4} {\bar D}^2 D_\a V~,
\quad V = {\bar V}~,
\ee
which involves an intrinsically real mass parameter.
The mass parameter is also intrinsically real 
in the vector-tensor  realization \cite{Siegel} 
(inspired by \cite{CS})
\bea
S_{\rm v-t} &=& - \hf \intss\, G^2 
+ {1\over 4}  \int {\rm d}^6z \,W^\a W_\a
+M \intss\, G \,V   
\non \\
&=& - \hf \intss\, G^2 +
{1\over 4}  \int {\rm d}^6z \,
W^\a W_\a
- \hf M \left\{  \int {\rm d}^6z \, W^\a \j_\a
+{\rm c.c.} \right\} ~,
\label{n=1v-t}
\eea
which describes the same  multiplet
on the mass shell
(massive superspin-1/2), and which 
possesses both the tensor multiplet gauge freedom 
\be
\d \j_\a = { {\rm i} \over 4} {\bar D}^2 D_\a K ~, 
\qquad K = {\bar K} 
\label{gauge-spinor}
\ee
and the vector multiplet gauge freedom 
\be
\d V = \L + {\bar \L}~, \qquad {\bar D}_\ad \L =0~.
\label{flat-vm-gauge}
\ee 

It was recently 
pointed out  \cite{DF,LS}, however,   
that giving the mass parameter
in (\ref{or-massive-tensor}) 
an imaginary part,\footnote{Unlike the scalar multiplet, 
this imaginary part cannot be eliminated by a rigid
phase transformation of $\j_\a$ as long as the 
explicit form of the linear scalar  $G$,  
in terms of $\j_\a$ and its conjugate, is fixed.}
\be
m^2 \quad \longrightarrow \quad 
m(m+{\rm i} e)~, 
\label{promotion}
\ee
leads to nontrivial physical implications,
for the mass can now be interpreted 
to have both electric and magnetic contributions
which are associated with the two possible 
mass terms $B \wedge {}^* B$ 
and $B \wedge B$ for the component two-form.
The dual vector multiplet is then characterized 
by the mass $M= \sqrt{m^2 +e^2}$.
What makes the replacement (\ref{promotion}) 
really interesting  is that  the complex mass parameter  
can be interpreted as a vacuum expectation value
for chiral scalars \cite{GL}, 
\be
m(m+{\rm i} e) \int {\rm d}^6z \, \j^\a \j_\a
\quad \longleftarrow \quad 
\int {\rm d}^6z \, F(\f) \,\j^\a \j_\a  ~, 
\ee
with $\f$ some   chiral scalars, 
${\bar D}_\ad \f =0$. 

In the massive case, the gauge freedom 
(\ref {gauge-spinor}) is broken if one does not 
use the vector-tensor formulation (\ref{n=1v-t}).
It can be restored, however, if  one 
implements the standard Stueckelberg
formalism, as was done in  \cite{DF,LS},
and replaces the naked chiral 
prepotential $\j_\a$ everywhere by 
\be 
\j_\a \quad \longrightarrow \quad 
\j_\a +  { {\rm i} \over m} W_\a~, 
\ee
where the compensating vector multiplet 
has to transform as  $\d V = m K $ under
(\ref{gauge-spinor}).
The gauge symmetry thus obtained can be treated
as a deformation of the transformations
(\ref{gauge-spinor}) and  (\ref{flat-vm-gauge}).

In this note we continue the research started in 
\cite{DF,LS} and provide further insight into 
the structure of massive tensor multiplets. 
In section 2 we consider aspects of $\cN=1$ tensor 
multiplets in curved superspace
and introduce a  model for the massive
improved tensor multiplet.  
Unlike the ordinary tensor
multiplet \cite{Siegel}, the (massless) improved 
tensor multipet \cite{deWR} is superconformal 
in global supersymmetry 
and super Weyl invariant
in curved superspace.
There are at least two reasons
why the improved tensor multiplet 
is interesting: (i) it describes 
the superconformal compensator in the new
minimal formulation of $\cN=1$ supergravity, 
see \cite{GGRS,BK} for reviews;
(ii) it corresponds to the Goldstone multiplet 
for partial breaking of $\cN=1$ superconformal 
symmetry associated with the coset 
$SU(2,2|1)/(SO(4,1) \times U(1)) $  
 which has  $AdS_5$ as the bosonic subspace
\cite{KM9,BIK2}.  As we demonstrate below, 
a remarkable feature of the improved tensor 
multiplet is that its super Weyl invariance 
remains intact in the massive case. 

In section 3 we introduce several realizations 
for the massive $\cN=2$ tensor multiplet, 
 describe its duality to the massive 
$\cN=2$ vector multiplet, and also sketch 
 possible self-couplings and couplings 
to vector multiplets. Section 4 is devoted 
to the description of the reduction of 
manifestly $\cN=2$ supersymmetric actions
to $\cN=1$ superspace. The list of $\cN=2$ 
superspace integrations measures is given 
in the appendix. 

Our $\cN=1$ supergravity conventions 
correspond to \cite{BK}. They are very similar 
to those adopted in \cite{WB}. 
The conversion from \cite{BK} to \cite{WB} 
is as follows: $E^{-1} \to E$ and $R \to 2R$.

\sect{$\cN=1$ tensor multiplets}
It is known that 4D $\cN=1$ 
new minimal supergravity
can be treated
as a super Weyl invariant dynamical system describing
the coupling of old minimal supergravity  to a {\it real}
covariantly linear scalar superfield ${\Bbb L}$
constrained by
$\Dsqac\, {\Bbb L} = (\cD^2 - 4 {\bar R})\, {\Bbb L} = 0$,
see \cite{GGRS,BK} for  reviews.\footnote{In 
old minimal supergravity, the superspace
covariant derivatives are 
$\cD_A =(\cD_a ,\cD_\a ,  {\bar \cD}^\ad ) 
=E_A{}^M (z)\,\pa_M + \O_A{}^{\b\g}(z)\, M_{\b\g} 
+\O_A{}^{\bd \gd}(z) \,{\bar M}_{\bd \gd}$, 
with $M_{\b\g}$ and ${\bar M}_{\bd \gd}$
the Lorentz generators. They obey the (modified) 
Wess-Zumino constraints, and the latter imply that 
the torsion and the curvature are expressed in terms
of  a vector $G_a = {\bar G}_a $ and covariantly chiral 
objects $R$ and $W_{\a \b \g}$  subject to some 
additional Bianchi identities.}
The new minimal supergravity action
\be\label{nmsg}
S_{\rm SG,new} = {3 \over \k^2} \intss\, E^{-1}\,
{\Bbb L}\, {\rm ln} {\Bbb L}~, \qquad \quad
E = {\rm Ber} (E_A{}^{M})
\ee
is invariant with respect to  
the super Weyl transformation\footnote{Under
(\ref{superweyl}), the full superspace measure 
changes as
${\rm d}^8 z\, E^{-1} \to
{\rm d}^8 z\, E^{-1} \,\exp (\s + \bar \s ) $,
while the {\it chiral superspace measure} transforms as
${\rm d}^8 z\, E^{-1} / R \to
{\rm d}^8 z\, (E^{-1} /R ) \, \exp (3\s )$.}  \cite{HT}
\be
\cD_\a ~\to~ {\rm e}^{ \s/2 - {\bar \s} } \Big(
\cD_\a - (\cD^\b \s) \, M_{\a \b} \Big) ~, \qquad
\cDB_\ad ~\to~ {\rm e}^{ {\bar \s}/2 - \s } \Big(
\cDB_\ad -  (\cDB^\bd {\bar \s}) {\bar M}_{\bd\ad} \Big)
\label{superweyl}
\ee
accompanied with 
\be
{\Bbb L} ~\to ~ {\rm e}^{-\s - \bar \s} \, {\Bbb L}~.
\ee
where $\s(z) $ is an 
arbitrary  covariantly chiral scalar parameter,
$\cDB_\ad \s=0$.
The super Weyl transformation of $\Bbb L$ is uniquely 
fixed by the constraint (\ref{linear}).
The dynamical system (\ref{nmsg}) is 
classically equivalent to old minimal supergravity 
described by the action 
\be\label{omsg}
S_{\rm SG,old} = - {3 \over \k^2} 
\intss\,E^{-1}~.
\ee

Modulo sign, the functional (\ref{nmsg}) 
coincides with the action for the so-called improved tensor 
multiplet \cite{deWR}
\be
\label{improved}
S = - \m \intss\, E^{-1}\,
G\, {\rm ln} (G / \m)~,
\ee
with $G$ obeying the same constraint 
as $\Bbb L$ above,
\be
\Dsqac\, G =
(\cD^2 - 4 {\bar R})\, G = 0~.
\label{linear}
\ee
In the family of tensor multiplet 
models \cite{Siegel} of the form
\be
S = \m^2 \intss\, E^{-1}\,
\cF (G / \m)~, 
\ee
the action (\ref{improved}) is singled out 
by the requirement of super Weyl invariance.
In particular, the free massless 
tensor multiplet action 
\be
S = - \hf \intss\, E^{-1}\,G^2 
\ee
is not super Weyl invariant.
Upon reduction to flat superspace, 
the action (\ref{improved}) 
becomes superconformal. 

As is well known, the general solution of 
(\ref{linear}) is 
\be
G = \hf ( \cD^\a \j_\a + {\bar \cD}_\ad {\bar \j}^\ad )~, 
\qquad  \quad {\bar \cD}_\ad \j_\a =0~, 
\label{strength}
\ee
with an arbitrary covariantly chiral spinor 
superfield $\j_\a$. The super Weyl transformation 
of the prepotential $\j_\a$ turns out to be uniquely fixed 
\cite{BK}:
\be
G ~\to ~ {\rm e}^{-\s - \bar \s} \,G \quad \Longrightarrow
\quad \j_\a ~\to~ {\rm e}^{-3\s/2} \, \j_\a~. 
\ee
As a result, adding a mass term to the 
action (\ref{improved}) does not spoil its 
super Weyl invariance. That is, the action 
\be
\label{mass-improved}
S [\j, \bar \j ]= - \m \intss\, E^{-1}\,
G\, {\rm ln} (G / \m)
-\hf m \left\{ (m+{\rm i}\,e) \intss\, {E^{-1} \over R}\, \j^2
+{\rm c.c.} \right\}
\ee
is invariant under arbitrary super Weyl transformations.
The latter property uniquely singles out this model 
 in the family of actions 
\be
\label{massive-tensor}
S = 
\m^2 \intss\, E^{-1}\,
\cF (G / \m)
- \hf m\left\{ (m+{\rm i}\,e) \intss\, {E^{-1} \over R}\, \j^2
+{\rm c.c.} \right\}~.
\label{general-model}
\ee
Therefore, the  action (\ref{mass-improved}) defines
the massive improved tensor multiplet.
This is a nontrivial theory, unlike the massless
improved tensor multiplet that  is known to be free.
Upon reduction to flat superspace, the action 
turns into a superconformal model.

Let us consider a dual formulation for the theory 
introduced in (\ref{mass-improved}).
We follow the procedure given in \cite{Siegel,GGRS,BK}
and first relax the linear constraints
 $\Dsqac\, G = (\cD^2 - 4 {\bar R})\, G = 0$ 
by introducing the ``first-order'' model
\bea
\label{mass-improved2}
S_{\rm auxiliary}  &=& - \m \intss\, E^{-1}\,
G\, \Big( {\rm ln} (G / \m) -1 \Big)
+ M \intss\, E^{-1}\, V \Big( 
G - \hf( \cD^\a \j_\a + {\bar \cD}_\ad {\bar \j}^\ad) \Big) \non \\
&&\qquad \qquad 
- \hf m\left\{ (m+{\rm i}\,e) \intss\, {E^{-1} \over R} \,\j^2
+{\rm c.c.} \right\}~, 
\eea
with 
\be
M^2 = m^2 +e^2~.
\label{massa}
\ee
Here both scalars $G$ and $V$ are real unconstrained, 
and $G$ is not related to $\j_\a$ and its conjugate. 
To preserve the super Weyl invariance, however,
$V$ should transform as follows:
\be
V ~\to ~ V - {\m  \over M} \,(\s + \bar \s ) ~.
\label{superweyl2}
\ee
Varying $S_{\rm auxiliary}$
 with respect to $V$ brings 
us back to (\ref{mass-improved}).
On the other hand,  varying $S_{\rm auxiliary}$ 
with respect to $G$ and $\j_\a$ allows one to express
these variables in terms of $V$ and 
the vector multiplet strength 
\be 
W_\a = -{1\over 4} ({\bar \cD}^2 -4R) \cD_\a V~,
\qquad {\bar \cD}_\ad W_\a =0~, 
\qquad \cD^\a W_\a = {\bar \cD}_\ad W^\ad ~.
\ee
One ends up with 
\be
S[V] = {1\over 4}  \intss\, {E^{-1} \over R} \,W^2
+\m^2 \intss\, E^{-1}\, \exp \Big( {M\over \m} V\Big)~.
\label{vector-improved} 
\ee
This action is invariant under the 
super Weyl transformations (\ref{superweyl})
and (\ref{superweyl2}).
It is worth pointing out that the inhomogeneous 
piece on the right of  (\ref{superweyl2})
does not show up in the transformation of 
$W_\a$: 
\be
W_\a ~\to~ {\rm e}^{-3\s/2} \, W_\a~. 
\ee
The super Weyl transformations of 
the chiral spinors $\j_\a$  and $W_\a$ are clearly 
identical.

Employing the Stueckelberg formalism, 
the action (\ref{vector-improved}) can be replaced
by the classically-equivalent one 
\be
S[V, \F , {\bar \F}] = {1\over 4}  \intss\, {E^{-1} \over R} \,W^2
+\m^2 \intss\, E^{-1}\, {\bar \F}\,{\rm e}^{(M/\m)V}
\F ~,
\label{vector-improved2} 
\ee
with $\F$ a compensating chiral scalar 
possessing a non-vanishing v.e.v. 
This action is  invariant under 
the $U(1)$ gauge transformation
\be
\d V = \L + {\bar \L}~, 
\qquad 
\d \F = - { \m \over M} \, \L\, \F~, 
\qquad {\bar \cD}_\ad \L=0~.
\label{n=1v-m-gauge}
\ee
The super Weyl transformation 
(\ref{superweyl2}) turns into 
\be
V ~\to ~ V ~, 
\qquad 
\F ~\to ~ {\rm e}^{-\s} \,\F~.
\label{superweyl3}
\ee
The model (\ref{vector-improved}), 
or its equivalent realization  
(\ref{vector-improved2}), 
describes the dynamics of a
massive improved vector multiplet
in curved superspace. 

The mass term in (\ref{mass-improved}) 
breaks the massless gauge symmetry  \cite{Siegel}
\be
\d \j_\a = { {\rm i} \over 4} ({\bar \cD}^2 -4R)\cD_\a K ~, 
\qquad K = {\bar K} 
\label{gauge-tra}
\ee
that leaves the field strength 
(\ref{strength}) invariant.  
Nevertheless, inspired by \cite{CS}, 
one can preserve the gauge symmetry  
in the massive case by 
considering the following vector-tensor model
\bea
S[\j, {\bar \j},V] =
- \m \intss\, E^{-1}\,
G\, {\rm ln} (G / \m)
+ {1\over 4}  \intss\, {E^{-1} \over R} \,W^2
+M \intss\, E^{-1}\,
G\, V~.
\label{v-t-supergr}
\eea
This action possesses both the tensor multiplet 
 (\ref{gauge-tra})  and  vector multiplet 
(\ref{n=1v-m-gauge}) gauge symmetries.
This action can also be seen to be super Weyl 
invariant provided   $V$ is chosen, say,  
to be inert under such transformations. 
By inspecting the equations of motion, 
one can check that the theory (\ref{v-t-supergr})
is classically equivalent to (\ref{mass-improved}) 
if  $M$ is chosen as in (\ref{massa}).
One can also establish the duality of 
(\ref{v-t-supergr}) to the improved vector
multiplet  (\ref{vector-improved2})
by dualizing the linear superfield $G$ 
into a chiral scalar and its conjugate
according to  \cite{Siegel,LR}.

In the massive case, following Stueckelberg,
the gauge  invariance (\ref{gauge-tra})
can be restored by introducing 
a compensating  Abelian vector multiplet 
(with the gauge field $\cV$ and 
the chiral  field strength $\cW_\a)$
and replacing $\j_\a$ in (\ref{general-model})
by 
\be 
\j_\a \quad \longrightarrow \quad 
\j_\a +  { {\rm i} \over m} \cW_\a~, 
\qquad \cW_\a = -{ 1 \over 4} 
({\bar \cD}^2 -4R) \cD_\a \cV ~, 
\quad \cV = {\bar \cV}~.
\ee
Here $\cV$ transforms 
as $\d \cV = m \,K$ under (\ref{gauge-tra}) 
such that the combination 
$m \, \j_\a +  {\rm i} \,\cW_\a $ is gauge invariant.
The modified mass term remains to be super Weyl 
invariant. Since 
$$
{\rm Im}  \intss\, {E^{-1} \over R} \,\cW^2 =0~,
$$
we then obtain 
\be
{\rm i}\,m  \intss\, {E^{-1} \over R} \, 
\Big(\j +  { {\rm i} \over m} \cW \Big)^2
+{\rm c.c.} = 
 {\rm i}  \intss\, {E^{-1} \over R} \, 
\Big(m \j^2 +  2 {\rm i} \, \j \cW \Big)
+{\rm c.c.} 
\ee

\sect{$\cN=2$ tensor multiplets}
To generalize the previous consideration to the
case of $\cN=2$ supersymmetry, it is 
advantageous (in some respect, necessary)
to make use 
of the $\cN=2$ harmonic superspace
${\Bbb R}^{4|8} \times S^2$
\cite{GIKOS,GIOS}.  
It extends conventional $\cN=2$ 
superspace ${\Bbb R}^{4|8} $ 
(paramerized by coordinates 
$Z= (x^a , \q^\a_i , {\bar \q}_\ad^i)$, 
with $i =\hat{1},  \hat{2}$) 
by the two-sphere $S^2 = SU(2)/U(1)$
parametrized by harmonics, i.e., group
elements
\bea
({u_i}^-\,,\,{u_i}^+) \in SU(2)~, \quad
u^+_i = \ve_{ij}u^{+j}~, \quad \overline{u^{+i}} = u^-_i~,
\quad u^{+i}u_i^- = 1 ~.
\eea
${}$For simplicity,
our consideration will be restricted 
to the study of globally  supersymmetric theories
only. 

Let us start by recalling the model for a free
massive $\cN=2$ vector multiplet \cite{GIKOS,GIOS}.
Its dynamical variable 
$V^{++} (Z,u)$ is a real analytic superfield, 
$D^+_\a V^{++}={\bar D}^+_\ad V^{++} =0$, 
where the harmonic-dependent spinor covariant 
derivatives  $D^\pm_\a$ and  
${\bar D}^\pm_\ad $
are defined in eq. (\ref{spinor-der}).
The action\footnote{The various $\cN=2$ 
superspace integration measures are defined
in the Appendix.}  is
\bea
S_{\rm vector} &=& \hf \int 
{\rm d}^8Z\, W^2 - \hf M^2 
\int {\rm d}\zeta^{(-4)}\,
(V^{++})^2 
\non \\
&=& \hf \int
{\rm d}^{12}Z\,{\rm d}u {\rm d}u'  \,
\frac{V^{++}( u)\,V^{++}(u')}
{(u^+u'^+)^2} 
- \hf M^2  \int {\rm d}\zeta^{(-4)}\,
(V^{++})^2~,
\label{n=2vector}
\eea
see \cite{GIOS} for the definition of harmonic 
distributions of the form $(u^+_1 u^+_2)^{-n}$, 
where $(u^+_1u^+_2) = u^{+}_1{}^i u^+_{2}{}_i$.
Here $W(Z)$ is the (harmonic independent) 
chiral field strength of the $\cN=2$  vector multiplet
\cite{GSW}, 
\be
{\bar D}^i_\ad W =0~, \qquad 
D^{\a i} D_\a^{j} W 
= {\bar D}_\ad^{i} {\bar D}^{j \ad} {\bar W}~, 
\ee
which is expressed via 
the analytic prepotential $V^{++} (Z,u)$
as follows \cite{GIOS,Z}:
\bea
W(Z)= {1\over 4} \int {\rm d}u \, 
({\bar D}^-)^2 \,V^{++}(Z,u)
&=&{1\over 4} ({\bar D}^+)^2 \,V^{--}(Z,u)~, 
\label{n=2vmfs} \\
V^{--} (Z,u) &=& \int {\rm d} u' \, 
\frac{V^{++}(Z, u')}
{(u^+u'^+)^2} ~.
\non
\eea
The equation of motion is
\be
{1\over 4} (D^+)^2 W - M^2\, V^{++} =0~,
\ee
where one should keep in mind that 
the Bianchi identity is equivalent
 to $ (D^+)^2\, W= ({\bar D}^+)^2 {\bar W}$.
This equation implies that $V^{++}$ is an $\cN=2$ 
linear superfield: 
\be 
D^{++} V^{++} =0 \quad  \longrightarrow \quad
V^{++}(Z,u) = V^{(ij)} (Z) u^+_i u^+_j ~, 
\label{lin1}
\ee
where $V^{ij} $ obeys the constaints 
\be
D^{(i}_\a V^{jk)} = {\bar D}^{(i}_\ad V^{jk)} =0
\quad  \longleftarrow \quad
D^+_\a V^{++} = {\bar D}^+_\ad V^{++}=0~, 
\label{lin2}
\ee
as a consequence of the analyticity
of the dynamical variable.
It is now easy to arrive at 
\be
(\Box -M^2)V^{++} =0
 \quad  \longrightarrow \quad
(\Box -M^2)W =0~.
\ee

In the massless case, $M=0$, 
the action (\ref{n=2vector}) is invariant 
under the gauge transformation 
\cite{GIKOS,GIOS}
\be 
\d V^{++} = D^{++} \l ~, 
\label{N=2vector-gauge-tr}
\ee
with the gauge parameter $\l(Z,u) $ a real 
analytic superfield,
$D^+_\a \l={\bar D}^+_\ad \l =0$. 
This transformation leaves the field strength
(\ref{n=2vmfs}) invariant.

Let us now turn to the massless $\cN=2$ 
tensor multiplet \cite{N=2tensor} formulated 
in harmonic superspace in \cite{GIO1,GIOS}. 
The free  action is
\bea
S &=&  \hf 
\int {\rm d}\zeta^{(-4)}\,(G^{++})^2~,
\label{N=2massless-tensor}
\eea
where $G^{++}(Z,u)$ is 
a restricted real analytic superfield
under the constraints 
(\ref{lin1}) and (\ref{lin2}). One can  
can express 
$G^{++}(Z,u) =G^{ij}(Z) \,u^+_iu^+_j$
in terms of an unconstrained 
chiral superfield $\J(Z)$
and its conjugate:
\be
G^{++}(Z,u) = {1\over 8} (D^+)^2\, \J (Z)
+{1\over 8} ({\bar D}^+)^2\, {\bar \J} (Z) ~, 
\qquad {\bar D}^i_\ad \J =0~. 
\ee
This superfield remains invariant under 
the gauge transformation
\be
\d \J = {\rm i} \, \L ~, \qquad
 {\bar D}_\ad^i \L =0~, \qquad 
D^{\a i} D_\a^{j} \L 
= {\bar D}_\ad^{i} {\bar D}^{j \ad} {\bar \L}~.
\label{N=2tensor-gauge-tr}
\ee
As is seen,  the chiral gauge parameter
$\L$ satisfies the same constraints 
as the vector multiplet field strength.

${}$Recalling the construction of  \cite{CS}, 
the massive tensor  (or vector) multiplet
can be described by the action 
\bea
S_{\rm v-t}&=&  \hf 
\int {\rm d}\zeta^{(-4)}\,(G^{++})^2
+ \hf \int 
{\rm d}^8Z\, W^2 
+M \int {\rm d}\zeta^{(-4)}\,
G^{++}\,V^{++}  \label{n=2v-t1}\\
&=&  \hf 
\int {\rm d}\zeta^{(-4)}\,(G^{++})^2
+ \hf \int 
{\rm d}^8Z\, W^2 
+\hf M \left\{  \int {\rm d}^8Z \, W \J
+{\rm c.c.} \right\} ~,
\eea
which is invariant under the gauge 
transformations (\ref{N=2vector-gauge-tr})
and (\ref{N=2tensor-gauge-tr}).
The corresponding equations of motion are
\be
{1 \over 4} ({\bar D}^-)^2 G^{++} 
+M \,W =0~, 
\qquad 
{1 \over 4} (D^+)^2 W + M G^{++} =0~,
\ee
as well as  the complex conjugate 
of the first equation.

Of primary importance for us will be
another  massive extension of 
(\ref{N=2massless-tensor}) 
\bea
S_{\rm tensor} &=&  \hf 
\int {\rm d}\zeta^{(-4)}\,(G^{++})^2
-{1\over 4} m\Big\{  (m+{\rm i}e) \int 
{\rm d}^8Z\, \J^2 
+{\rm c.c.} \Big\} ~. 
\label{n=2tensor}
\eea
This action generates the following 
equations of motion 
\be
{1 \over 4} ({\bar D}^-)^2 G^{++} 
-m (m+{\rm i}e) \, \J=0~, 
\qquad 
{1 \over 4} (D^-)^2 G^{++} 
-m (m-{\rm i}e) \, {\bar \J}=0~,
\ee
which imply 
\be 
(\Box -M^2) G^{++} =0~, 
\qquad M = \sqrt{m^2 +e^2}~.
\ee

The dynamical systems  (\ref{n=2vector})
and (\ref{n=2tensor}) turn out to  be 
dual to each other 
provided $M$ is chosen as above.
The corresponding ``first-order'' action, 
which establishes the duality between  these
theories, is 
\bea
S_{\rm auxiliary} =  \hf 
\int {\rm d}\zeta^{(-4)}\,(G^{++})^2
&+& {1\over 8} M \int {\rm d}\zeta^{(-4)}\,V^{++}
\Big( 8G^{++} -  (D^+)^2\, \J 
-({\bar D}^+)^2\, {\bar \J} \Big)  
\nonumber \\
&- &{1\over 4}m \Big\{  (m+{\rm i}e) \int 
{\rm d}^8Z\, \J^2 
+{\rm c.c.} \Big\} ~,
\eea
where both real analytic superfields
$V^{++}$ and $G^{++}$ are unconstrained.
Varying $V^{++}$ brings us back to 
(\ref{n=2tensor}). 
On the other hand, varying 
$S_{\rm auxiliary} $ with respect to 
$G^{++}$ and $\J$ and using the equations 
obtained to eliminate these superfields, 
we end up with (\ref{n=2vector}). 

One can also establish duality between 
(\ref{n=2vector}) and the gauge-invariant 
model (\ref{n=2v-t1}) by 
using the known duality between 
the massless tensor multipet and the 
$\o$-hypermultiplet \cite{GIO1}.
Consider the  ``first-order'' action 
\bea
\tilde{S}_{\rm v-t}=   
\int {\rm d}\zeta^{(-4)}\,
\left\{ \hf (G^{++})^2 +M G^{++}\,V^{++}
+ G^{++} \,D^{++}\o \right\}
+ \hf \int  {\rm d}^8Z\, W^2 ~, 
\eea
in which $G^{++}$ and $\o$ are real 
unrestricted analytic superfields. 
Varying $\o$ gives $D^{++} G^{++} = 0$, 
and then (\ref{n=2v-t1})  is reproduced.
On the other hand, varying $G^{++}$ 
and then eliminating it from $\tilde{S}_{\rm v-t}$, 
we arrive at the action
\be 
\hat{S}_{\rm vector}
= \hf \int  {\rm d}^8Z\, W^2 
-\hf M^2 \int {\rm d}\zeta^{(-4)}\,
\Big( V^{++} + {1 \over M}\,D^{++}\o \Big)^2~,
\ee
and this is simply the Stueckelberg{\it ization} of 
(\ref{n=2vector}).

In the massive case, the gauge freedom
(\ref{N=2tensor-gauge-tr}) is broken.  
It can be restored, following Stueckelberg, 
by introducing a compensating  
Abelian vector multiplet 
(with the gauge field $\cV^{++}$ and 
the chiral  field strength $\cW$) and 
replacing the naked prepotential 
$\J$ as  follows:
\be
\J \quad \longrightarrow \quad \J+ 
{ {\rm i} \over m} \cW~,  \qquad 
\d \cW = - m \L~.
\ee
The combination $m\J+ {\rm i} \, \cW$
is invariant under the gauge
transformations  (\ref{N=2tensor-gauge-tr}).  
One then obtains
\bea
m \, {\rm i} \int 
{\rm d}^8Z\, 
\Big(\J+  { {\rm i} \over m} \cW \Big)^2 
+{\rm c.c.}
= {\rm i} \int  {\rm d}^8Z\, 
\Big(m\, \J^2+  2 {\rm i} \, \J\,\cW \Big) 
+{\rm c.c.}
\eea
Up to this point, the use of $\cN=2$ harmonic
superspace has allowed us to keep 
a complete analogy with 
the $\cN=1$ case previously considered. 

Let us turn to possible generalizations
to generate (self-)interactions. 
A natural extension\footnote{One can also 
add two supersymmetric gauge-invariant terms 
${\rm Re}\Big\{c_1 \int {\rm d}\zeta^{(-4)}\,G^{++}  (\q^+)^2 
+ c_2 \int {\rm d}^8Z\, \J  \Big\}$, with 
$c_{1,2}$ complex parameters, 
which trigger spontaneous 
supersymmetry breaking.} 
 of the kinetic term 
(\ref{N=2massless-tensor}) 
is \cite{GIO1,GIOS}
\bea 
\hf \int {\rm d}\zeta^{(-4)}\,(G^{++})^2
\quad & \longrightarrow &  \quad 
S_{\rm H} =
\int {\rm d}\zeta^{(-4)}\,\cL^{(+4)} (G^{++}, u^+, u^-)~. 
\label{general-kin}
\eea
Here $\cL^{(+4)} $ is an arbitrary (real analytic) 
function of the field strength $G^{++}$ and the harmonics
$u^\pm$ carrying $U(1) $ charge $+4$. 
In particular, the improved $\cN=2$ tensor multiplet 
\cite{deWPV,LR,KLR,GIO1} 
is generated by \cite{GIO1,GIOS}
\bea
\cL^{(+4)} _{\rm impr}(G^{++}, u) = \m^2 \,
\Big( { \cG^{++}  \over 1 + \sqrt{ 1+ \cG^{++} \,c^{--}  }}
\Big)^2~, \qquad 
\cG^{++} = G^{++}/ \m - c^{++} ~,
\eea 
with $c^{++}$ a holomorphic vector field on $S^2$, 
\be
c^{\pm \pm }(u) = c^{ij} \,u^\pm_i u^\pm_j ~, \qquad 
c^{ij} c_{ij} =2~, \qquad c^{ij} = {\rm const}~.
\ee
The corresponding action takes a simpler form
in the so-called projective superspace 
\cite{KLR,LR2}.

We can introduce the massive improved 
tensor multiplet
\bea
S_{\rm impr} =
\int {\rm d}\zeta^{(-4)}\,
\cL^{(+4)}_{\rm impr} (G^{++}, u)
- {1\over 4} m\Big\{  (m+{\rm i}e) \int 
{\rm d}^8Z\, \J^2 
+{\rm c.c.} \Big\} ~.
\label{n=2-impr}
\eea
Here the kinetic term is known to be invariant 
under the $\cN=2$ superconformal group 
\cite{GIO1,GIOS}.
The mass term turns out to be 
superconformally invariant as well. 
In fact, the above action is the most general 
superconformal action without higher derivatives.
Instead of (\ref{n=2-impr}), we  can work
with the gauge-invariant action
\bea
S'_{\rm impr}&=&  
\int {\rm d}\zeta^{(-4)}\,
\cL^{(+4)}_{\rm impr} (G^{++}, u)
+ \hf \int 
{\rm d}^8Z\, W^2 
+M \int {\rm d}\zeta^{(-4)}\,
G^{++}\,V^{++}  ~,
\label{n=2-impr2}
\eea
which respects the gauge symmetries
 (\ref{N=2vector-gauge-tr})
and (\ref{N=2tensor-gauge-tr}).
The linear superfield $G^{++}$ in 
(\ref{n=2-impr2}) can be further dualized into
a real analytic superfield $\o$ ($\o$-hypermultiplet), 
thus converting the action (\ref{n=2-impr2}) 
into that for a  massive improved vector multiplet.

The mass term in (\ref{n=2tensor}) 
admits a natural generalization  of 
the form
\bea
m (m+{\rm i}e) \int 
{\rm d}^8Z\, \J^2 
\quad & \longrightarrow &  \quad 
\int {\rm d}^8Z\, \U (\J, {\Bbb W} ) ~, 
\label{holomorphic-mass}
\eea
where $\U$ is a holomorphic function, and 
$\Bbb W$ stands for the chiral field strength(s) 
of some vector multiplet(s).
Unlike the $\cN=1$ supersymmetric case, 
where the chiral prepotential $\j_\a$ was 
anticommuting, here we have no inherent 
reasons to insist that  $\U (\J, {\Bbb W} )$ 
be quadratic in $\J$.

\sect{From $\cN=2$ superfields to 
$\cN=1$ superfields}

Here we describe the reduction of 
$\cN=2$ tensor multiplet models to $\cN=1$
superspace. The latter is parametrized 
by the coordinates
$z= (x^a , \q^\a , {\bar \q}_\ad)$ 
related to the $\cN=2$ superspace 
coordinates $Z= (x^a , \q^\a_i , {\bar \q}_\ad^i)$, 
with $i =\hat{1},  \hat{2}$, 
as follows:
\be
 \q^\a  = \q_{\hat{1}}{}^\a ~,  \qquad 
{\bar \q}_\ad = {\bar \q}^{\hat{1}}{}_\ad~.
\ee
The $\cN=1$ spinor covariant derivatives
$(D_\a , {\bar D}^\ad)$ are related 
to the $\cN=2$ 
covariant derivatives $(D_\a^i , {\bar D}^\ad_i)$ 
in a similar fashion, 
\be
 D_\a  = D^{\hat{1}}{}_\a ~,  \qquad 
{\bar D}^\ad = {\bar D}_{\hat{1}}{}^\ad~.
\ee
${}$For any $\cN=2$ superfield $U(Z) 
=U(x , \q_i , {\bar \q}^j)$, 
we define its $\cN=1$ projection
\be
U| = U(z) = U(x , \q_i , {\bar \q}^j)
\Big|_{
\q_{\hat{2}} =  {\bar \q}^{\hat{2}}  =0}~. 
\ee

Start with the $\cN=2$ tensor multiplet strength
\be
G^{ij} = G^{ji}
= {1\over 8} D^i D^j\, \J 
+{1\over 8} {\bar D}^i {\bar D}^j\, {\bar \J}  ~, 
\qquad {\bar D}^i_\ad \J =0~,
\ee
which obeys the constraints
\be
D^{(i}_\a G^{jk)} = {\bar D}^{(i}_\ad G^{jk)} =0~.
\ee
The prepotential $\J(Z)$ 
reduces to the three $\cN=1$ chiral components:
\be
\s = \J|~, \qquad 
{\rm i}\,\j_\a = \hf  D^{\hat{2}}{}_\a \,\J|~, \qquad 
\r =-{1 \over 4} D^{\hat{2}}D^{\hat{2}}\,\J|~.
\ee
Then, for the components of $G^{ij}$ we get
\be
G^{ \hat{2} \hat{2}} | \equiv \F 
= -\hf \Big( \r  - {1\over 4}  {\bar D}^2 {\bar \s} \Big)~,
\qquad G^{ \hat{2} \hat{1}} | ={ {\rm i}  \over 2} \,G ~,
\qquad 
G^{ \hat{1} \hat{1}} | 
= {\bar \F} ~,
\ee
with $G$ the $\cN=1$ tensor multiplet strength, 
eq. (\ref{ten-field-strength}). 

Now, for the $\cN=2$ chiral mass-like  term 
(\ref{holomorphic-mass}) with ${\Bbb W} =0$ 
we obtain
\be
\int {\rm d}^8Z\, \U (\J) 
= -  \int {\rm d}^8z\,{\bar \s}^I \, \U_I (\s  ) 
+ \int {\rm d}^6z\, \Big\{ \j^I \j^J \, \U_{IJ}(\s) 
-2\F^I\,\U_I(\s) \Big\}~,
\label{holom-couping1}
\ee
where we have specialized to the 
case of several tensor multiplets.
This is, of course, very similar to the $\cN=1$ 
form of the holomorphic  prepotential in the 
Seiberg-Witten theory.

To reduce the free kinetic term
for the tensor multiplet, eq.  
(\ref{N=2massless-tensor}),
to $\cN=1$ superspace,
one can use the identity \cite{BK2}
\be
 \int {\rm d}\zeta^{(-4)}\,F^{++} \,G^{++}
=  \int {\rm d}^8z\,\Big\{ 
F^{ \hat{1} \hat{1}} | \,G^{ \hat{2} \hat{2}} | 
+F^{ \hat{2} \hat{2}} | \, G^{ \hat{1} \hat{1}} | 
+4F^{ \hat{1} \hat{2}} | \,G^{ \hat{1} \hat{2}} | 
\Big\}~,
\ee
with $F^{++}$ a linear superfield
(i.e. an analytic superfield under 
the same constraints that $G^{++}$ obeys). 
This gives
\be
\hf  \int {\rm d}\zeta^{(-4)}\,(G^{++})^2
= \int {\rm d}^8z\,\Big\{ 
{\bar \F} \F - \hf \,G^2 \Big \}~. 
\ee

More general kinetic terms for the tensor multiplet, 
eq. (\ref{general-kin}),
are easier to analyze using the projective superspace 
techniques \cite{KLR,LR2}.
In projective superspace, 
the $\cN=2$ supersymmetric action 
involves integration over a closed loop in 
the complex plane ${\Bbb C}$, 
unlike the harmonic superspace
action (\ref{general-kin}) that involves
integration over $S^2$.
The projective-superspace action 
for the $\cN=2$ tensor multiplet
\cite{KLR,LR2}  is 
\be 
S_{\rm P}=
\frac{1}{2\p {\rm i}} \oint_\g
\frac{ {\rm d} w}{w} \int {\rm d}^8z\,
\cL ( \S(w), w ) 
+{\rm c.c.}
\label{projective-action}
\ee
Here $\cL$ is an
arbitrary  ``good'' function, 
$\g$ an appropriately chosen contour, and 
\be
\S (w) = \F + w \,G - w^2 \,{\bar \F}~.
\ee
The action (\ref{projective-action}) can be
shown to be $\cN=2$ supersymmetric, 
in spite of the fact that it involves integration 
only over the $\cN=1$ superspace.
The relationship between the harmonic
action $S_{\rm H}$ in  (\ref{general-kin}) 
and the projective action (\ref{projective-action}) 
was studied in \cite{K}.

As an example, let us consider the special choice
\cite{GHK} that corresponds to the so-called 
$c$-map \cite{CFG}: 
$\cL ( \S, w ) =  \Xi (\S) / w^2$, 
with $\Xi$ a holomorphic function. 
In addition, the  contour $\g$ in 
(\ref {projective-action}) should now
enclose  the origin. Then we obtain 
\bea
 \frac{1}{2\p {\rm i}} \oint_\g
\frac{ {\rm d} w}{w} \int {\rm d}^8z\,
{ \Xi( \S(w) ) \over w^2} 
= -  \int {\rm d}^8z\,{\bar \F}^I \, \Xi_I (\F  ) 
+ \hf \int {\rm d}^8z\,  G^I G^J \, \Xi_{IJ}(\F) ~,
\label{holom-couping2}
\eea
where we have specialized to the 
case of several tensor multiplets.
Comparing the first terms on the right of
(\ref{holom-couping1}) and (\ref{holom-couping2}), 
we see that they are of the same functional form.
We therefore believe that  $\cN=2$ 
supersymmetric theories of the form  
\be
S = \frac{1}{2\p {\rm i}} \oint_\g
\frac{ {\rm d} w}{w} \int {\rm d}^8z\,
{ \Xi ( \S(w) ) \over w^2} 
+\int {\rm d}^8Z\, \U (\J, {\Bbb W})  + {\rm c.c.}
\ee
that are  generated by two holomorphic potentials,
$\Xi$ and $\U$,  deserve further study.

In the main body of this note, we studied 
models for a single massive tensor multiplet
in $\cN=1$ and $\cN=2$ supersymmetry. 
The results can be clearly extended to 
the case of several mutiplets. 

Shortly  before the submission of this
note to the hep-th  archive, 
we received a new paper \cite{Theis}
in which the supersymmetric 
Freedman-Townsend models
\cite{LR,CLL} (see also \cite{FINN})
and their generalizations
\cite{BT} were made massive
by extending the non-supersymmetric 
construction of \cite{DFTV}.

\vskip.5cm
\noindent
{\bf Acknowledgements:}\\ 
Discussions with Ian McArthur 
are gratefully acknowledged.
This work is supported in part by the Australian Research
Council.

\begin{appendix}

\sect{$\cN=2$ superspace integration measures}
Here we introduce  various $\cN=2$ superspace 
integration measures used throughout this paper. 
They are defined
in terms of the spinor covariant derivatives
$D^i_\a$ and ${\bar D}^\ad_i$, with 
$i =\hat{1},  \hat{2}$, 
\be
\{ D^i_\a , D^j_\b \} 
= \{ {\bar D}^\ad_i , {\bar D}^\bd_j \} =0~, 
\qquad 
\{ D^i_\a , {\bar D}_{j \bd} \} = -2{\rm i}\,
\d^i_j \,(\s^m)_{\a \bd} \, \pa_m~,
\ee
and the related fourth-order operators
\bea
 D^4 = {1\over 16} (D^{\hat{1}} )^2 (D^{\hat{2}})^2 
~,  \qquad 
{\bar D}^4 = {1\over 16} 
({\bar D}_{\hat{1}} )^2 ({\bar D}_{\hat{2}})^2 ~.
\eea
Integration over the chiral subspace is defined by 
\be 
\int {\rm d}^8 Z \, L_{\rm c} = 
\int {\rm d}^4 x\, D^4 L_{\rm c}~, 
\qquad {\bar D}^i_\ad L_{\rm c} =0~.
\ee
Integration over the full superspace 
is defined by 
\be 
\int {\rm d}^{12} Z \, L = 
\int {\rm d}^4 x\, {\bar D}^4 D^4 L~. 
\ee
In terms of the harmonic-dependent 
spinor derivatives 
\be
D^\pm_\a = D^i_\a \,u^\pm_i~, \qquad 
{\bar D}^\pm_\ad = {\bar D}^i_\ad \,u^\pm_i~,
\label{spinor-der}
\ee 
and the related fourth-order operators 
\be 
(D^+)^4 = {1 \over 16} (D^+)^2 ({\bar D}^+)^2 ~, 
\qquad  
(D^-)^4 = {1 \over 16} (D^-)^2 ({\bar D}^-)^2 ~, 
\ee
integration over the analytic subspace is defined by
\be 
\int {\rm d}\z^{(-4)} \, L^{(+4)} = 
\int {\rm d}^4 x \int {\rm d}u\, (D^-)^4 L^{(+4)}~, 
\qquad 
D^+_\a L^{(+4)} =
{\bar D}^+_\ad L^{(+4)} =0~.
\ee
Integration over the group manifold 
$SU(2)$ is defined according to \cite{GIKOS}
\be 
 \int {\rm d}u \, 1 = 1~\qquad 
 \int {\rm d}u \, u^+_{(i_1} \cdots u^+_{i_n}\,
u^-_{j_1} \cdots u^-_{j_m)} =0~, 
\quad n+m >0~.
\ee

\end{appendix}

\small{

}

\end{document}